\documentclass[aps,prd,twocolumn,showpacs,superscriptaddress,longbibliography,nofootinbib,preprintnumbers,amsmath,amssymb,floatfix,10pt]{revtex4-1}

\usepackage{amsmath,amssymb}
\usepackage{natbib}
\usepackage{hyperref}
\usepackage{bbm}
\usepackage{graphicx}
\usepackage{xcolor}
\usepackage{mathrsfs}

\newcommand{\bw}{\begin{widetext}}
\newcommand{\ew}{\end{widetext}}

\newcommand{\be}{\begin{equation}}
\newcommand{\ee}{\end{equation}}
\newcommand{\bestar}{\begin{equation*}}
\newcommand{\eestar}{\end{equation*}}

\newcommand{\bi}{\begin{itemize}}
\newcommand{\ei}{\end{itemize}}
\newcommand{\bea}{\begin{eqnarray}}
\newcommand{\eea}{\end{eqnarray}}

\newcommand{\hbo}{\hbox to 1 true cm {\hfill } }
\newcommand{\tr}{\hbox{tr}\,}

\newcommand{\sotimes}{{\, \scriptstyle \otimes \, }}

\newcommand{\sfrac}[2]{{\textstyle \frac{#1}{#2}}}

\newcommand{\Eins}{\mathbbmss{1}}
\newcommand{\vc}[1]{\mbox{\boldmath$#1$}}

\newcommand{\vcb}[1]{\mbox{\bf #1}}

\newcommand{\m}{{\scriptscriptstyle -}}
\newcommand{\p}{{\scriptscriptstyle +}}

\def\lambdabar{\protect\@lambdabar}
\def\@lambdabar{%
\relax \bgroup
\def\@tempa{\hbox{\raise.73\ht0
\hbox to0pt{\kern.2\wd0\vrule width.7\wd0
height.1pt depth.1pt\hss}\box0}}%
\mathchoice{\setbox0\hbox{$\displaystyle\lambda$}\@tempa}%
{\setbox0\hbox{$\textstyle\lambda$}\@tempa}%
{\setbox0\hbox{$\scriptstyle\lambda$}\@tempa}%
{\setbox0\hbox{$\scriptscriptstyle\lambda$}\@tempa}%
\egroup }

\begin{document}
\title{Signatures of Radiation Reaction in Ultra-Intense Laser Fields}

\author{Chris Harvey}
\affiliation{Department of Physics, Ume\aa\ University, SE-901 87 Ume\aa, Sweden}

\author{Thomas Heinzl}\email[]{theinzl@plymouth.ac.uk}
\affiliation{School of Computing and Mathematics, University of Plymouth, Plymouth PL4 8AA, UK}

\author{Mattias Marklund}
\affiliation{Department of Physics, Ume\aa\ University, SE-901 87 Ume\aa, Sweden}

\begin{abstract}
\noindent We discuss radiation reaction effects on charges propagating in ultra-intense laser fields. Our analysis is based on an analytic solution of the Landau-Lifshitz equation. We suggest to measure radiation reaction in terms of a symmetry breaking parameter associated with the violation of null translation invariance in the direction opposite to the laser beam. As the Landau-Lifshitz equation is nonlinear the energy transfer within the pulse is rather sensitive to initial conditions. This is elucidated by comparing colliding and fixed target modes in electron laser collisions.
\end{abstract}

\maketitle
\section{Introduction}
The problem of classical radiation reaction (RR) has vexed generations of physicists since its first formulation in 1892 by Lorentz \cite{Lorentz:1892,Lorentz:1909}. Following important contributions by Abraham \cite{Abraham:1905} and others, the equation describing the back reaction of the radiation field on the motion of the radiating charge has been cast in its final covariant form by Dirac in 1938 \cite{Dirac:1938}. It is now aptly called the Lorentz-Abraham-Dirac (LAD) equation. The relevant body of literature has become enormous and we refer to the recent monographs \cite{Spohn:2004ik,Rohrlich:2007} and, in particular, to the preprint \cite{McDonald:1998} for an overview of the historical development and extensive lists of references.

A particularly compact way of writing the LAD equation, say for an electron (mass $m$, charge $e$) is
\be \label{LAD}
  m \dot{u} = F + \tau_0 \, \mathbbm{P} \, m \ddot{u} \; ,
\ee
where $u$ denotes the electron 4-velocity,  $F = e \mathbbm{F}u/c$ the Lorentz 4-force in terms of the field strength tensor, $\mathbbm{F}$, of the \emph{externally prescribed} field and dots derivatives with respect to proper time, $\tau$. The second term on the right is the RR force, $F_\mathrm{RR}$, which is characterised by the appearance of the time parameter
\be
  \tau_0 = \sfrac{2}{3} r_e/c \simeq 2 \; \mbox{fm}/c \simeq  10^{-23} \; \mbox{s} \; .
\ee
This is the time it takes light to traverse the classical electron radius\footnote{We employ Heaviside-Lorentz units with fine structure constant $\alpha = e^2 /4\pi\hbar c = 1/137$.},
\be
  r_e = e^2/4\pi mc^2 = \alpha \lambdabar_C \simeq 3 \, \mbox{fm} \; ,
\ee
or, from the second term, the electron Compton wave length reduced by a factor $\alpha \simeq 1/137$, the fine structure constant. Obviously, the time and length scales involved are typical for higher order QED corrections (or even strong interactions, i.e.\ QCD) -- a hint that the classical LAD  equation (\ref{LAD}) may not capture all the physics at these microscopic (and essentially quantum) scales. Finally, the projection $\mathbbm{P} \equiv \Eins - u \sotimes u/c^2$ in (\ref{LAD}) guarantees that 4-acceleration and velocity are Minkowski orthogonal\footnote{We denote the tensor product $u_\mu u_\nu$ in index-free notation with the standard symbol, ``$\sotimes$''.}. This follows upon differentiating the on-shell condition, $u^2 = c^2$, which, of course, is Einstein's postulate on the universality of the speed of light, $c$. As the velocity $u$ is time-like, the acceleration $\dot{u}$ is space-like, $\dot{u}^2 < 0$.

\section{Estimating Radiation Reaction}

The LAD equation (\ref{LAD}) is of third order in time derivatives ($\ddot{u} = \dddot{x}$) and hence suffers from a number of pathologies such as runaway solutions and pre-acceleration. One way to overcome this is by iteration, assuming that $F_{\mathrm{RR}} \ll F$ -- which amounts to working to first order in $\tau_0$. This in turn implies a `reduction of order' in derivatives and results in the Landau-Lifshitz (LL) equation \cite{Landau:1987},
\be \label{LL}
  m \dot{u} = F + \tau_0 \, \mathbbm{P} \, \dot{F} \; .
\ee
Hence, one replaces the offending `jerk' \cite{Russo:2009yd} term, $m \ddot{u}$, in (\ref{LAD}) by the proper time derivative of the Lorentz force \cite{Rohrlich:2007} where the $\dot{u}$ term is evaluated to \emph{lowest} order in $\tau_0$, giving
\be \label{DOT.F}
  \dot{F} = \frac{e}{c} \, \dot{\mathbbm{F}}u + \frac{e^2}{mc^2} \, \mathbbm{F}^2 u + O(\tau_0) \; .
\ee
For alternative derivations of the LL equation resolving mathematical intricacies related to regularisations of the point particle concept we refer to  \cite{Spohn:1999uf,Gralla:2009md}.

The LL equation (\ref{LL}) was derived under the assumption of a small reaction force, $F_{\mathrm{RR}} \ll F$. Let us elucidate the physics involved somewhat further by assuming that the external field is produced by a laser described by a plane wave with light-like wave vector $k$, $k^2 = 0$. An electron `approaching' the laser field with initial 4-velocity $u_0$ will, in its rest frame, `see' a wave frequency given by the scalar product,
\be
  \Omega_0 \equiv k \cdot u_0  \; .
\ee
At this point one has to distinguish between two points of view. If both $k$ and $u_0$ are simultaneously Lorentz transformed the frequency $\Omega_0$ remains, of course, invariant. On the other hand, one may think of $k$, the wave vector of the laser as measured in the lab, as a distinguished 4-vector that breaks explicit Lorentz invariance (selecting a specific photon energy and beam direction). Different choices of initial conditions (i.e.\ $u_0$) then characterise the relation between the initial rest frame of the electron and the lab frame. In what follows we will adopt this second point of view.

The temporal gradients in (\ref{DOT.F}) will be of the order of the laser period, $\dot F \simeq \Omega_0 F$, so that the relative magnitude of the reaction force becomes
\be \label{RR1}
  r \equiv \frac{F_\mathrm{RR}}{F} = \Omega_0 \tau_0 \ll 1 \; ,
\ee
with the inequality required for the validity of the LL approximation. Consider now a head-on collision in the lab where the laser wave vector and electron velocity are given by
\be \label{K.U}
  k = \omega/c \, (1, \hat{\vcb{z}}) \; , \quad  u_0 = \gamma_0 c \, (1, - \beta_0 \hat{\vcb{z}}) \; ,
\ee
with the usual relativistic gamma factor, $\gamma_0 = E_e/mc^2$ measuring the electron energy $E_e$ in units of $mc^2$. Such an electron then `sees' a laser frequency that is Doppler upshifted according to
\be
  \Omega_0 = \gamma_0 (1 + \beta_0) \omega \equiv e^{\zeta_0} \omega \simeq 2\gamma_0 \omega \; ,
\ee
with the last identity holding for $\gamma_0 \gg 1$. This boost in laser frequency is just the usual energy gain due to colliding versus fixed target mode (which, of course, are related by a longitudinal Lorentz boost with rapidity $\zeta_0$). If we define dimensionless photon energies in the co-moving and lab frames,
\be
  \nu_0 \equiv \frac{\hbar \Omega_0}{mc^2} \; , \quad \nu \equiv \frac{\hbar \omega}{mc^2} \; ,
\ee
the RR parameter $r$ from (\ref{RR1}) becomes
\be \label{RR2}
  r = \sfrac{2}{3} \alpha \nu_0 \simeq \sfrac{4}{3} \alpha \gamma_0 \nu \simeq 10^{-2} \gamma_0 \nu \; .
\ee
For an optical laser $\nu \simeq 10^{-6}$ so that $r \simeq 10^{-8} \gamma_0$. Thus, to boost this to order unity (such that reaction equals Lorentz force) requires $\gamma_0 \simeq 10^8$, i.e.\ electron energies of order $10^2$ TeV. These can only  be produced in gamma-ray bursts, but not (currently) in labs. The ground breaking laser pair production (``matter from light'') experiment SLAC E-144, for instance, was utilising the 50 GeV SLAC linear collider implying $\gamma_0 \simeq 10^5$ and $r \simeq 10^{-3}$ \cite{Bamber:1999zt}.

The standard way of quantifying radiation by accelerated charges is via Larmor's formula for the radiated power, the relativistic incarnation of which may be written as
\be \label{LARMOR}
  P = - m \tau_0 \dot{u}^2 > 0 \; ,
\ee
and hence is of order $\tau_0$. The LL equation (\ref{LL}) would thus give us the radiated power to order $\tau_0^2$. If we follow our philosophy of neglecting second-order terms it suffices to express the acceleration via just the Lorentz force,
\be \label{LORENTZ}
  m \dot{u} = F = e \mathbbm{F} u/c \equiv e E \; ,
\ee
where we have introduced the space-like 4-vector $E$ corresponding to the electric field `seen' by the electron. If we now assume transversality of our laser fields, $k \cdot E = 0$, we immediately derive a very useful conservation law by dotting $k$ into (\ref{LORENTZ}),
\be
  \dot{\Omega} \equiv k \cdot \dot{u} = 0 \; .
\ee
In other words, without RR, the electron always `sees' the \emph{same} laser frequency on its passage through the laser beam,
\be \label{CONSERV}
  \Omega = k \cdot u = k \cdot u_0 = \Omega_0 \; .
\ee
If we now plug the Lorentz equation (\ref{LORENTZ}) into (\ref{LARMOR}) the average energy loss per (reduced) laser period $1/\Omega_0$ in units of $mc^2$ becomes
\be \label{R}
  R \equiv \frac{\langle P \rangle }{\Omega_0 mc^2} = - r \, \frac{e^2 \langle E^2 \rangle}{m^2 c^2 \Omega_0^2} \equiv  r a_0^2 \; .
\ee
Interestingly, we recover our RR parameter $r$ from (\ref{RR1}) and (\ref{RR2}) multiplying a new quantity, the dimensionless laser amplitude $a_0$. This measures the energy gain of an electron traversing a laser wavelength, $\lambdabar_L = c/\Omega_0$, in a field of average strength $\langle -E^2\rangle^{1/2}$ in units of $mc^2$. Obviously, when this (purely classical) parameter becomes of order unity, the electron motion is relativistic. Note that $a_0$ is Lorentz invariant if $k$ and $u_0$ are transformed simultaneously. Gauge invariance is shown \cite{Heinzl:2008rh} by expressing it in terms of the field strength, $\mathbb{F}$ which is conveniently rendered dimensionless by introducing
\be \label{F.DIMLESS}
  \hat{\mathbbm{F}} \equiv e\mathbbm{F}/mc\Omega_0 \; ,
\ee
so that $a_0^2$ finally becomes
\be \label{A01}
  a_0^2 = \frac{(u_0, \langle \hat{\mathbbm{F}}^2 \rangle u_0)}{c^2} \; .
\ee
The energy loss parameter (\ref{R}) was previously employed in \cite{Koga:2005,DiPiazza:2008lm,DiPiazza:2009pk,Hadad:2010mt}. It suggests that for substantial radiation the small RR parameter $r$ needs to be compensated by large values of $a_0^2$. The magnitude of $a_0$ is most easily estimated by introducing Sauter's critical electrical field, $E_S \equiv m^2c^3/e\hbar = 1.3 \times 10^{18}$ V/m \cite{Sauter:1931zz} and the associated intensity, $I_S \equiv c E_S^2 = 4.2 \times 10^{29}$ W/cm$^2$. For a given lab intensity $I$  we then have
\be
  a_0 = \frac{1}{\nu} \, \sqrt{\frac{I}{I_{S}}} \simeq 10^6 \, \sqrt{\frac{I}{I_{S}}} \; .
\ee
In view of the current record intensity of $I = 10^{22}$ W/cm$^2$ \cite{Yanovsky:2008} one can envisage $a_0$ values of about $10^3$ for the not too distant future \cite{Vulcan10PW:2009,ELI:2009}. We have seen already in (\ref{RR2}) that large gamma factors (colliding mode) yield a further increase of radiative losses. In addition, the losses accumulate over successive laser periods. After, say, $N$ cycles one expects a total relative change of the electron gamma factor given by
\be \label{DELTA.GAMMA}
  \frac{|\Delta \gamma|}{\gamma_0} = 2\pi N R = \frac{8\pi}{3} N \alpha \gamma_0 \nu a_0^2 \; .
\ee
Thus, the smallness of $\alpha\nu \simeq 10^{-8}$ may be compensated by pulse duration, $N$, initial electron energy, $\gamma_0$, and intensity, $a_0^2$.

\section{Modelling the laser}

The simplest model of a laser (beam) is provided by a plane wave with a field tensor depending solely on the phase, $\mathbbm{F} = \mathbbm{F}(\phi)$, $\phi = k \cdot x$, and obeying transversality, $\mathbbm{F}k = 0$. If we choose $k$ as in (\ref{K.U}) we have
\bea
  \Omega &=& k \cdot u = \omega (u^0 - u^3)/c \equiv \omega u^\m/c \; , \label{U.MINUS} \\
  \phi &=& k \cdot x = \omega (x^0 - x^3)/c \equiv \omega x^\m / c \; . \label{X.MINUS}
\eea
In other words, the laser field $\mathbbm{F}$ only depends on the light-front or null coordinate, $x^\m = ct - z$ \cite{Heinzl:2000ht}.

Plane waves are invariantly characterised as \emph{null} fields \cite{Synge:1935zzb,Stephani:2004ud} for which both scalar and pseudoscalar invariants vanish,
\bea
  \mathscr{S} \equiv \sfrac{1}{4} \tr \mathbbm{F}^2 &=& 0 \; , \label{S} \\
  \mathscr{P} \equiv \sfrac{1}{4} \tr \mathbbm{F}\tilde{\mathbbm{F}} &=& 0 \; . \label{P}
\eea
The vanishing of $\mathscr{S}$ implies that the energy momentum tensor of a plane wave is just $\mathbbm{F}^2$,
\be \label{T}
  c\mathbbm{T} = \mathbbm{F}^2 - \mathscr{S}\Eins = \mathbbm{F}^2 \; .
\ee
This is the \emph{only} nontrivial power of field strength as $\mathbbm{F}$ is nilpotent with index 3, i.e.\ $\mathbbm{F}^3 = 0$, which will be important when we solve the equations of motion in such a field.

We emphasize that there is no \emph{intrinsic} invariant scale associated with a null field. The only way to associate a nonvanishing invariant is by using a probe (dubbed ``third agent'' in \cite{Becker:1991}) such as an electron or a (non-laser) photon. This naturally leads to the invariant amplitude $a_0$ as defined in (\ref{A01}) which explicitly depends on the probe electron 4-velocity. Defining the energy density of the laser `seen' by the electron as
\be \label{W0}
  w_0 \equiv (u_0, c\mathbbm{T}u_0)/c^2 = (u_0, \mathbbm{F}^2 u_0)/c^2 = E^2 \; ,
\ee
we see that (\ref{A01}) precisely represents the dimensionless version of this energy density,
\be \label{A02}
  a_0^2 = \langle \hat{w}_0 \rangle  \; .
\ee
Typically, the plane wave modelling the laser will be pulsed, i.e.\ of finite  duration in phase $\phi$. We accommodate this situation by parameterising the dimensionless field strength $\hat{\mathbbm{F}}$ as follows. We assume the plane wave field to be linearly polarised along the space-like transverse 4-vector $\epsilon$ ($\epsilon^2 = -1$) and hence decompose $\hat{\mathbbm{F}}$ into magnitude $a_0$, envelope $f= f(\phi)$ and a constant tensor, $\mathbbm{f} = n \sotimes \epsilon - \epsilon \sotimes n$,
\be \label{F}
  \hat{\mathbbm{F}}(\phi) = a_0 \, f(\phi) \, \mathbbm{f} \; ,
\ee
with the dimensionless constant 4-vector $n \equiv kc/\Omega_0$ obeying $n^2 = n \cdot \epsilon = 0$ and $n \cdot u = n \cdot u_0 = c$. As a result, the square of $\mathbbm{f}$ is simply $\mathbbm{f}^2 = n \sotimes n$ with all higher powers vanishing due to $n^2 = 0$. In order for (\ref{F}) to be consistent with (\ref{A01}) and $(\ref{A02})$ the average $\langle \ldots \rangle$ must be defined in such a way that $f$ is normalised, $\langle f^2 \rangle = 1$. Defining a dimensionless gauge potential, $\hat{A} = eA/mc^2$, the field strength becomes
\be \label{A}
  \hat{\mathbbm{F}} = n \sotimes \hat{A}' - \hat{A}' \sotimes n \; ,
\ee
the prime henceforth denoting the derivative with respect to invariant phase $\phi$. Comparison with (\ref{F}) finally yields $\hat{A}' = a_0 f \epsilon$ and $a_0^2 = - \langle \hat{A}^{\prime 2} \rangle$.

A more realistic laser model is provided by Gaussian beams, i.e.\ solutions of the wave equation in paraxial approximation \cite{Davis:1979zz}. The corresponding fields have nontrivial longitudinal and transverse envelopes resulting in the appearance of longitudinal field components. For this reason, the fields are no longer null and the charge dynamics becomes more complicated. Charged particle velocities and trajectories in such fields have to be obtained numerically \cite{Harvey:2011}.

\section{Solving the LL equation}

Unlike the LAD the LL equation is a fairly standard equation of motion being second order in time derivatives. Hence, it requires two integrations and initial conditions for velocity and position. For the purposes of the present discussion it will be sufficient to perform only the first integration for which we need to provide the initial 4-velocity, $u(0) = u_0$.

\subsection{Neglecting Radiation Reaction}

To set the stage for a later comparison we first briefly recall the solution \emph{without} RR, i.e.\ of the Lorentz force equation of motion (\ref{LORENTZ}). We note that for any function $f = f(\tau)$ of proper time we may trade derivatives according to
\be
  \dot{f} = f' \dot{\phi} = f' \Omega = f' \Omega_0 \; ,
\ee
where, in the last step, we have used the conservation law (\ref{CONSERV}). In terms of the dimensionless field strength $\hat{\mathbbm{F}}$ the Lorentz equation takes on the particularly compact form
\be \label{U.PRIME}
  u' = \hat{\mathbbm{F}} u \; .
\ee
This is, of course, integrated by a matrix exponential which truncates at second order due to nilpotency, $\hat{\mathbbm{F}}^3 = 0$. Employing the parameterisation (\ref{F}) the solution becomes
\be \label{LORENTZ.SOLUTION}
  u_L = u_0 - c a_0 I_1 \epsilon + \left( a_0 I_1 \epsilon \cdot u_0 + \sfrac{1}{2} a_0^2 I_1^2 c \right)\, n \; ,
\ee
where the subscript ``L'' stands for ``Lorentz''. The function $I_1 (\phi)$ is the pulse shape integral
\be \label{I.1}
  I_1 \equiv \int\limits_0^\phi d\varphi \, f (\varphi) \; .
\ee
Upon inspection of the solution (\ref{LORENTZ.SOLUTION}) we note the following features. The velocity decomposes into transverse and longitudinal contributions given by the second and third terms on the right, respectively. If the initial velocity is longitudinal (like for a head-on collision) we have $\epsilon \cdot u_0 = 0$. In this case, the longitudinal velocity is quadratic in $a_0$ while the transverse component is always linear.

One may rewrite the solution (\ref{LORENTZ.SOLUTION}) in terms of the gauge potential $\hat{A} = a_0 I_1 \epsilon$ defined in (\ref{A}) which results in the neat expression
\be \label{UA}
  u = u_0 - c \hat{A} + (\hat{A} \cdot u_0 - \sfrac{1}{2} \hat{A}^2 c) n
\ee
From this expression one easily identifies the additional conserved quantity $\epsilon \cdot (u + c\hat{A})$ \cite{Lai:1980} corresponding to the transverse canonical momentum. As stated above, one has $\hat{A} \cdot u_0 = 0$ for both fixed target and colliding modes. The quadratic contributions in (\ref{UA}) are actually positive as $\hat{A}$ is space-like, $\hat{A}^2 < 0$.

\subsection{Including Radiation Reaction}

Upon including RR we have to solve the full LL equation (\ref{LL}) which we write in dimensionless notation as
\be \label{LL.DIMLESS}
  \hat{\Omega}u' = \hat{\mathbbm{F}} u + r (\hat{\Omega} \hat{\mathbbm{F}}' + \hat{\mathbbm{F}}^2 - \hat{w})u \; ,
\ee
with  $\hat{w} \equiv (u, \hat{\mathbbm{F}}^2u)/c^2$, cf.\ (\ref{W0}) and (\ref{A02}), and a normalised frequency,
\be
  \hat{\Omega} \equiv \Omega/\Omega_0 = k \cdot u/ k \cdot u_0 \; .
\ee
Clearly, the LL equation (\ref{LL.DIMLESS}) is nonlinear in the unknown $u$. Remarkably, though, there is an analytic solution for a plane wave background, $\mathbbm{F} = \mathbbm{F} (\phi)$ \cite{DiPiazza:2008lm,Hadad:2010mt,Harvey:2010ns}. Let us briefly review its main steps using our compact notation.

Recall that the Lorentz equation (\ref{LORENTZ}) entails the conservation law (\ref{CONSERV}) which is just $\hat{\Omega} = 1$. In contradistinction, a non-vanishing RR force, $F_\mathrm{RR} = \tau_0 \mathbbm{P} \dot{F}$, implies that $\hat{\Omega}$ is no longer conserved, but rather
\be \label{OMEGA.PRIME}
  \hat{\Omega}' = - r \hat{w}^2 = - R f^2 \hat{\Omega}^2 \; .
\ee
This is possibly the most significant new feature: In the presence of RR the electron will see a \emph{continuously changing} laser frequency during its passage through the pulse. Crucially, however, the equation (\ref{OMEGA.PRIME}) for the longitudinal velocity component completely decouples and, being first-order, can be solved by straightforward quadrature,
\be \label{OMEGA}
  \hat{\Omega} = \frac{1}{1 + RI_2} \simeq 1 - R I_2 \; ,
\ee
assuming the initial condition $\hat{\Omega}_0 = 1$ and defining the shape integral
\be \label{I.2}
  I_2 \equiv \int_0^\phi d\varphi \, f^2 (\varphi) \; ,
\ee
cf.\ (\ref{I.1}). It is worth noting that the RR parameter $R$ from (\ref{R}) appears at this stage. As $R \sim \tau_0$ we should actually use the ultimate expression in (\ref{OMEGA}) in keeping with our philosophy of neglecting terms of order $\tau_0^2$.

In any case we would like to point out that $\hat{\Omega}$ is a particularly nice signature for RR as it differs from unity only when a substantial amount of RR is present. In more physical terms, $\hat{\Omega} \ne 1$ signals \emph{symmetry breaking} in the following sense. Together with the longitudinal velocity $u^0 - u^3 \sim \Omega$, cf.\ (\ref{U.MINUS}), the longitudinal momentum, $p^\m = p^0 - p^3$, ceases to be conserved. As a result RR induces a breaking of translational invariance in the conjugate null direction\footnote{Recall the scalar product $p \cdot x = p^\p x^\m /2 + p^\m x^\p / 2 - \vc{p}_\perp \cdot \vc{x}_\perp$.}, $x^\p = x^0 + x^3$.

Let us continue with the LL equation and the remaining velocity components. The crucial technical trick \cite{DiPiazza:2008lm} is to introduce a new 4-velocity $v$ via
\be
  v \equiv \hat{\Omega}^{-1} u = (1 + R I_2) u \; ,
\ee
the longitudinal component of which is again conserved, $k \cdot v = \Omega_0$. Using (\ref{OMEGA.PRIME}) it is straightforward to see that the LL equation for $v$ simplifies to
\be \label{V.PRIME}
  v' = ( \hat{\Omega}^{-1} \hat{\mathbbm{F}}  + r \hat{\mathbbm{F}}' + r \hat{\Omega}^{-1} \hat{\mathbbm{F}}^2) v \; .
\ee
As we know $\hat{\Omega}^{-1}$ as a function of $\phi$ from (\ref{OMEGA}), the system (\ref{V.PRIME}) is indeed linear and easily solved via exponentiation. To arrive at the solution we use the parameterisation (\ref{F}) and discard all terms of order $\tau_0^2$ (or $r^2$). Noting that $u_0 = v_0$ the solution for $v$ may be then be written as a correction to the Lorentz solution (\ref{LORENTZ.SOLUTION}) in the following way,
\be \label{V.SOLUTION}
  v = u_L + r \Delta v + O(r^2) \; .
\ee
The RR term is explicitly given by
\be \label{DELTA.V}
  \Delta v =  - c a_0 I_{21} \epsilon + \left( a_0 I_{21} \epsilon \cdot u_0 + a_0^2 I_1 I_{21} c + a_0^2 I_2 c \right) n  \; ,
\ee
with the new shape integral,
\be \label{I.21}
    I_{21} \equiv f + a_0^2 \int_0^\phi d\varphi \, I_2 f \; .
\ee
Comparing (\ref{DELTA.V}) and (\ref{LORENTZ.SOLUTION}) one can identify precisely the same vector structure, the analogous longitudinal and transverse terms guaranteeing $k \cdot v = k \cdot v_0 = \Omega_0 = const$. Obviously, in the limit of no RR ($r \to 0$) one has $\hat{\Omega} \to 1$ and $\Delta v \to 0$ such that (\ref{LORENTZ.SOLUTION}) is readily recovered from (\ref{V.SOLUTION}) and (\ref{DELTA.V}).


As a final comment we note that the dependence on proper time $\tau$ is recovered by integrating (\ref{OMEGA}), which gives
\be
  \Omega_0 \tau = \phi + R \int_0^\phi d\varphi \, I_2 (\varphi)\; .
\ee
Hence, in proper time $\tau$, RR leads to a phase shift compared to the Lorentz solution \cite{Harvey:2010ns} as $\tau$ is no longer proportional to the phase, $\phi$.

\begin{figure}[!h]
\includegraphics[scale=0.4]{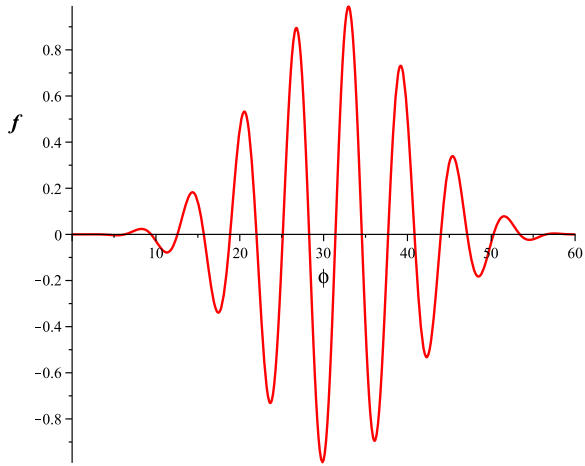}
\caption{\label{fig:PULSE}Laser pulse (\ref{PULSE}) for $N = 10$ as a function of phase, $\phi$.}
\end{figure}

\section{An analytic example}

With the analytic solution (\ref{DELTA.V}) of the LL equation at hand we can readily analyse an example. The only remaining technical difficulty is the evaluation of the pulse shape integrals (\ref{I.1}), (\ref{I.2}) and (\ref{I.21}). It turns out that they may be performed analytically for the pulse shape function
\be \label{PULSE}
  f (\phi) \equiv \left \{ \begin{array}{ll} \sin^4(\phi/2 N)\sin(\phi) \; , \quad &  0 \le \phi \le 2 \pi N  \; , \\
                                                                 0 \quad &  \mbox{else}
                                 \end{array} \right.
\ee
originally suggested in \cite{Mackenroth:2010jk} (see also \cite{Heinzl:2010vg}).  The pulse (\ref{PULSE}) has a duration of $\phi_0 \equiv 2 \pi N$ (hence contains $N$ cycles) and vanishes identically outside this interval. Thus, unlike a sine modulated Gaussian \cite{Harvey:2010ns}, it has compact support (see Fig.~\ref{fig:PULSE}).

\subsection{Symmetry breaking}

\begin{figure}
\includegraphics[scale=0.43]{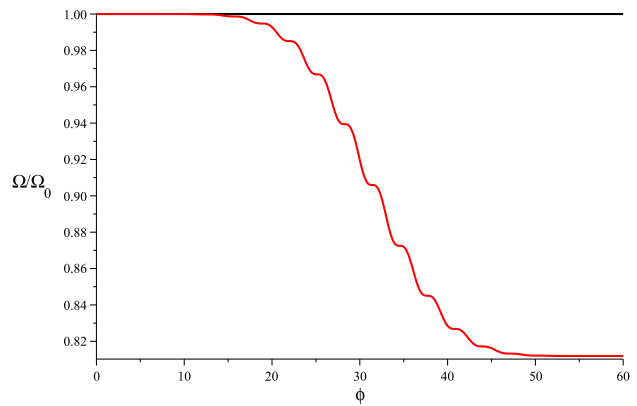}
\caption{\label{fig:OMEGA}Laser frequency $\hat{\Omega} = k \cdot u/ k \cdot u_0$ as `seen' by the electron during the passing of the pulse ($N=10$) as a function of phase, $\phi$. Horizontal (black) line: Constant result (\ref{CONSERV}) without RR (no symmetry breaking). Decreasing (red) line: RR Solution (\ref{OMEGA}) signalling symmetry breaking.
}
\end{figure}

Let us first consider the behaviour of the symmetry breaking parameter $\hat{\Omega}$ from (\ref{OMEGA}). For this we need the integral $I_2$, which is
\be \label{I2P}
  I_2 = \left \{ \begin{array}{ll} \frac{35}{256} \phi + \delta I_2 \quad & \quad 0 \le \phi \le \phi_0  \; , \\[5pt]
                                   \frac{35}{256} \phi_0 \quad & \quad \mbox{else}
                 \end{array} \right.     \; .
\ee
The term $\delta I_2$ in (\ref{I2P}) denotes a series of small amplitude sine functions which we do not explicitly display. All we need to know is that they vanish at the `end' of the pulse, $\delta I_2(\phi_0) = 0$. Hence, inside the pulse $\hat{\Omega} = 1 - R I_2$ drops linearly with small oscillations superimposed until it reaches a final plateau. For parameter values $\gamma_0 = 100$, $a_0 = 150$, $\nu = 10^{-6}$ (implying $R = 0.022$) and $N=10$ the resulting behaviour is shown in Fig.~\ref{fig:OMEGA}.

From (\ref{I2P}) the final plateau value, once the pulse has passed, is given by the simple expression
\be \label{OMEGA.F}
  \hat{\Omega}_f = 1 - R I_2 (\phi_0) = 1 - 2 \pi N \frac{35}{256} R  .
\ee
For the parameter values of Fig.~\ref{fig:OMEGA} the numerical value for the plateau value is $\hat{\Omega}_f = 0.81$. In general, assuming a head-on collision with $\gamma_0 \gg 1$ one has $\Omega_0 \simeq 2 \gamma_0$, $\Omega_f \simeq 2 \gamma_f$ and the total relative energy loss becomes
\be \label{DELTA.GAMMA.LL}
  \frac{|\Delta \gamma|}{\gamma_0} \simeq 1 - \hat{\Omega}_f =   \frac{35}{256} \times 2 \pi N R \simeq NR \; .
\ee
Apart from the numerical coefficient this is precisely our prediction (\ref{DELTA.GAMMA}) based on Larmor's formula.

\begin{figure}
\includegraphics[scale=0.43]{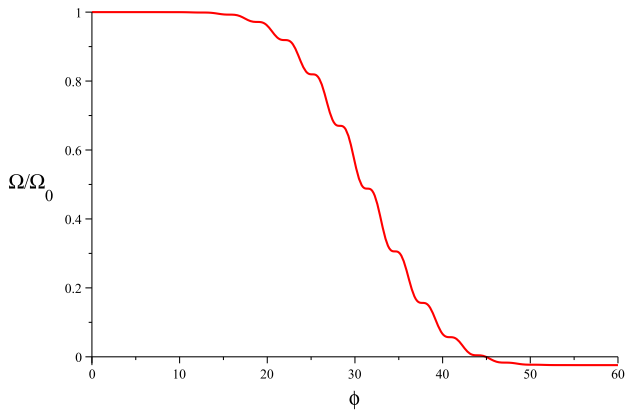}
\caption{\label{fig:OMEGA.CRIT}Laser frequency $\hat{\Omega}$ as a function of phase, $\phi$, for a pulse with $N=10$ and $a_0 = 350$. For this value, $\hat{\Omega}$ becomes negative signalling a breakdown of the LL approximation.
}
\end{figure}

As an additional bonus (\ref{OMEGA.F}) provides us with a criterion for when the LL approximation breaks down. This clearly is the case when $NR$ becomes of order unity which, using (\ref{RR2}), translates into
\be
   e^{\zeta_0} a_0^2 N \simeq 10^8 \; .
\ee
For $\gamma_0 \gg 1$ one has $e^{\zeta_0} = \gamma_0 (1+ \beta_0) \simeq 2 \gamma_0$, so, when $\gamma_0 = 100$ in Fig.~\ref{fig:OMEGA} we expect our approximations to break down when $a_0^2 \simeq 10^5$ or $a_0 \simeq 320$. This is indeed borne out by Fig.~\ref{fig:OMEGA.CRIT}.

\subsection{Varying initial conditions}

Let us finally look at the energy transfer dynamics in more detail. We want particularly to compare the two scenarios of fixed target and colliding modes which can both be described by the choice (\ref{K.U}). All we have to do for a fixed target (electron initially at rest) is set $\beta_0 = 0$ and $\gamma_0 = 1$. We are interested in the electron energy as a function of $\phi$, which describes its ``history'' during the passing of the pulse. This energy may be written as
\be
  E_e = mc u^0 = \gamma m c^2 \; ,
\ee
so we just have to monitor the behaviour of the electron gamma factor, $\gamma (\phi) = u^0 (\phi)/c$. In the LL case this is governed by a nonlinear system, so we expect a significant dependence on initial conditions. This is indeed what happens. Let us first work out the analytical expressions.

\begin{figure}
\includegraphics[scale=0.43]{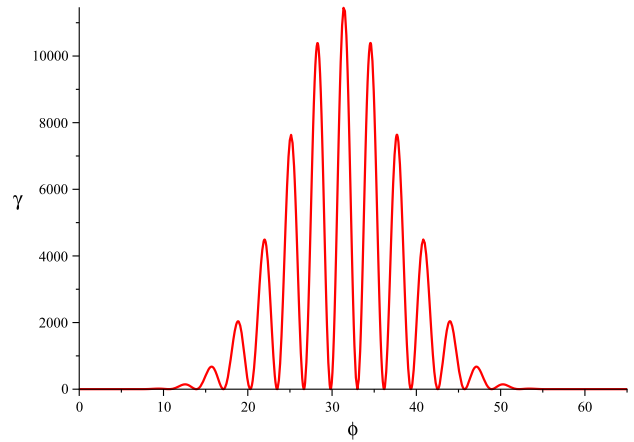}
\caption{\label{fig:GAMMA1} Electron gamma factor as a function of phase $\phi$ in fixed target mode ($\gamma_0 = 1$). At the given vertical scale the curves with and without RR are indistinguishable. In both cases there is no net energy transfer.}
\end{figure}

In analogy with (\ref{V.SOLUTION}) we split into a Lorentz and RR part,
\be
  \gamma = \gamma_L + \Delta \gamma \; ,
\ee
with the Lorentz contribution (\ref{LORENTZ.SOLUTION}) always yielding an increase,
\be
  \gamma_L = \gamma_0 + \frac{1}{2} e^{-\zeta_0} a_0^2 \, I_1^2 > \gamma_0 \; .
\ee
The leading RR correction is
\be \label{DELTA.GAMMA.I}
  \Delta \gamma = \frac{2}{3} \alpha \nu a_0^2 \Big\{I_1 f + a_0^2 I_4 + I_2 \Big(1 - \gamma_0 e^{\zeta_0} \Big)  \Big \} \; ,
\ee
with the new combination of shape integrals,
\be
  I_4 \equiv I_1 \int_0^\phi I_2 f - \frac{1}{2} I_1^2 I_2 \; .
\ee
For the pulse (\ref{PULSE}) $I_4$ turns out to be positive with compact support while $I_1 f$ can have either sign. All shape integrals are of order unity. Hence, for the case of interest (large $a_0 \gg 1$) the positive $I_4$ term dominates the $I_1 f$ term. As a consequence, the sign of $\Delta \gamma$ is entirely determined by the last term, that is, by the initial conditions! The simplest case is the fixed target mode (FTM, $\gamma_0 = 1 = e^{\zeta_0}$) for which the crucial term vanishes and the RR correction is never negative,
\be \label{DELTA.FTM}
  \Delta \gamma_{\mathrm{FTM}} \simeq \frac{2}{3} \alpha \nu \, a_0^4 \, I_4 \ge 0 \; .
\ee
Asymptotically, once the pulse has passed ($\phi > \phi_0$) we have $\Delta \gamma_{\mathrm{FTM}} = 0$ as $I_4$ has compact support. Hence, for FTM (and \emph{only} for FTM!) there is no net energy transfer between electron and laser pulse (Fig.~\ref{fig:GAMMA1}).

\begin{figure}
\includegraphics[scale=0.43]{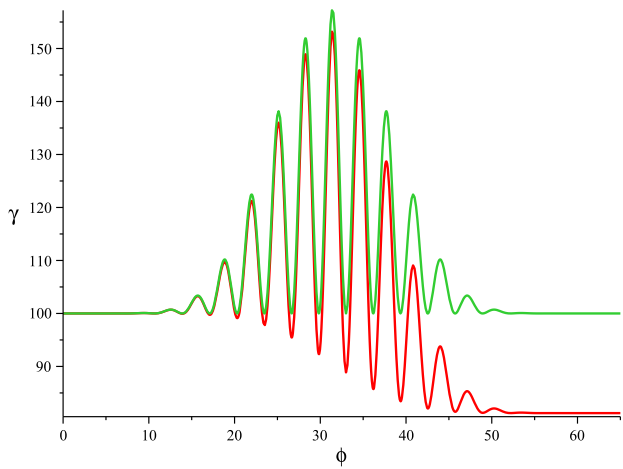}
\caption{\label{fig:GAMMA.L.VS.LL} Electron gamma factor as a function of phase $\phi$ in colliding mode ($\gamma_0 = 100$). Upper curve (green): Lorentz equation (no RR). Lower curve (red): LL equation (RR) which implies net energy loss.}
\end{figure}

The situation is different in colliding mode (CM). Assuming $\gamma_0 \gg 1$ we have
\be
  \Delta \gamma_{\mathrm{CM}} \simeq \frac{2}{3} \alpha \nu a_0^2 \left( a_0^2 I_4 - 2 \gamma_0^2 I_2 \right) \; .
\ee
Unlike $I_4$, $I_2$ takes on a nonzero constant value after the pulse ($\phi > \phi_0$). Hence, there is an asymptotic energy \emph{loss},
\be
  \Delta \gamma_{\mathrm{CM}}/\gamma_0  \simeq  - \frac{4}{3} \alpha \nu a_0^2 \gamma_0 I_2 (\phi_0) \simeq - NR \; ,
\ee
so that we recover (\ref{DELTA.GAMMA.LL}) having identified its sign. The behaviour of $\gamma$ in CM is depicted in Fig.~\ref{fig:GAMMA.L.VS.LL} both with and without RR.

\begin{figure}
\includegraphics[width=0.45\textwidth]{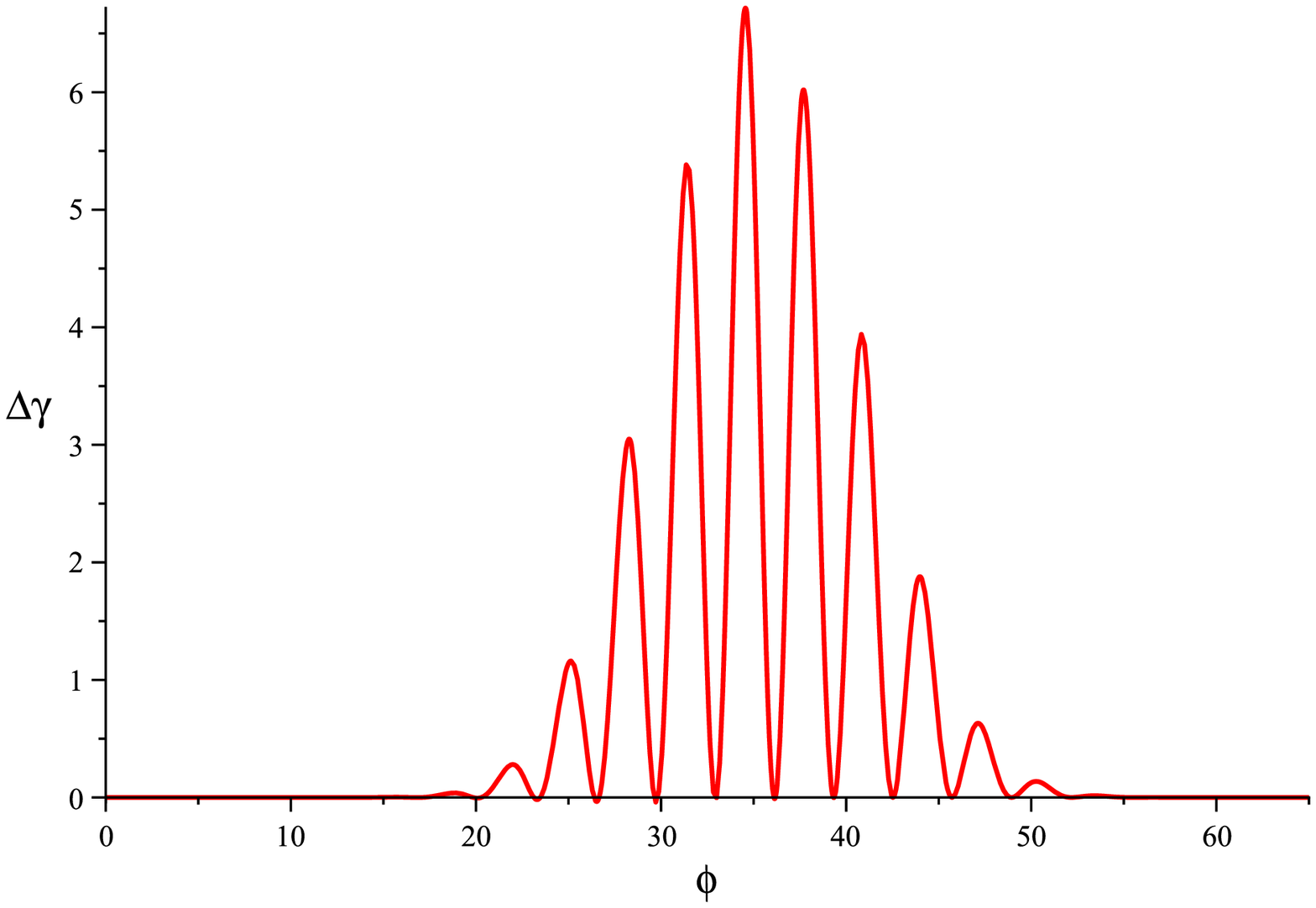}
\includegraphics[width=0.47\textwidth]{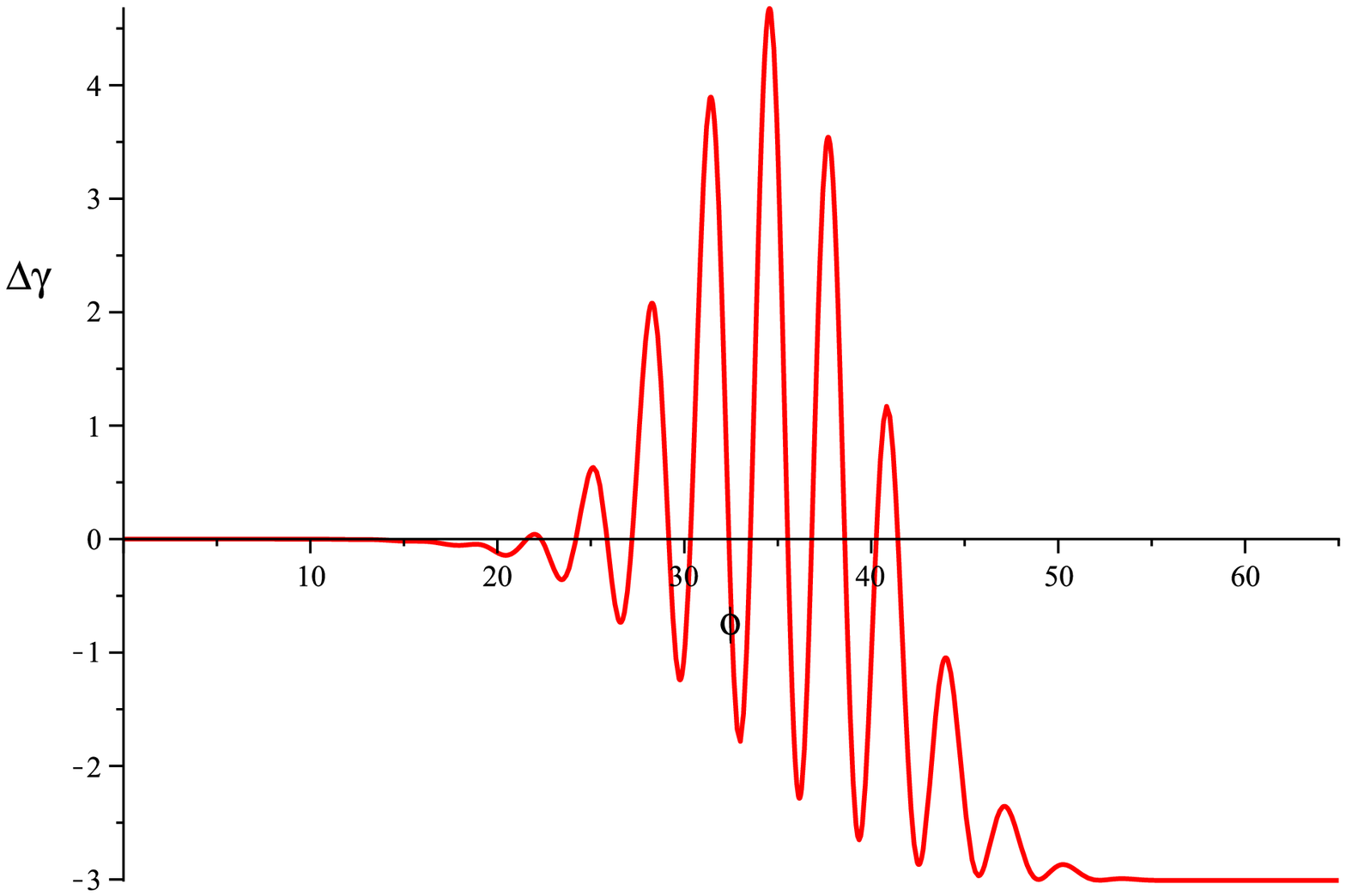}\\[5pt]
\includegraphics[width=0.48\textwidth]{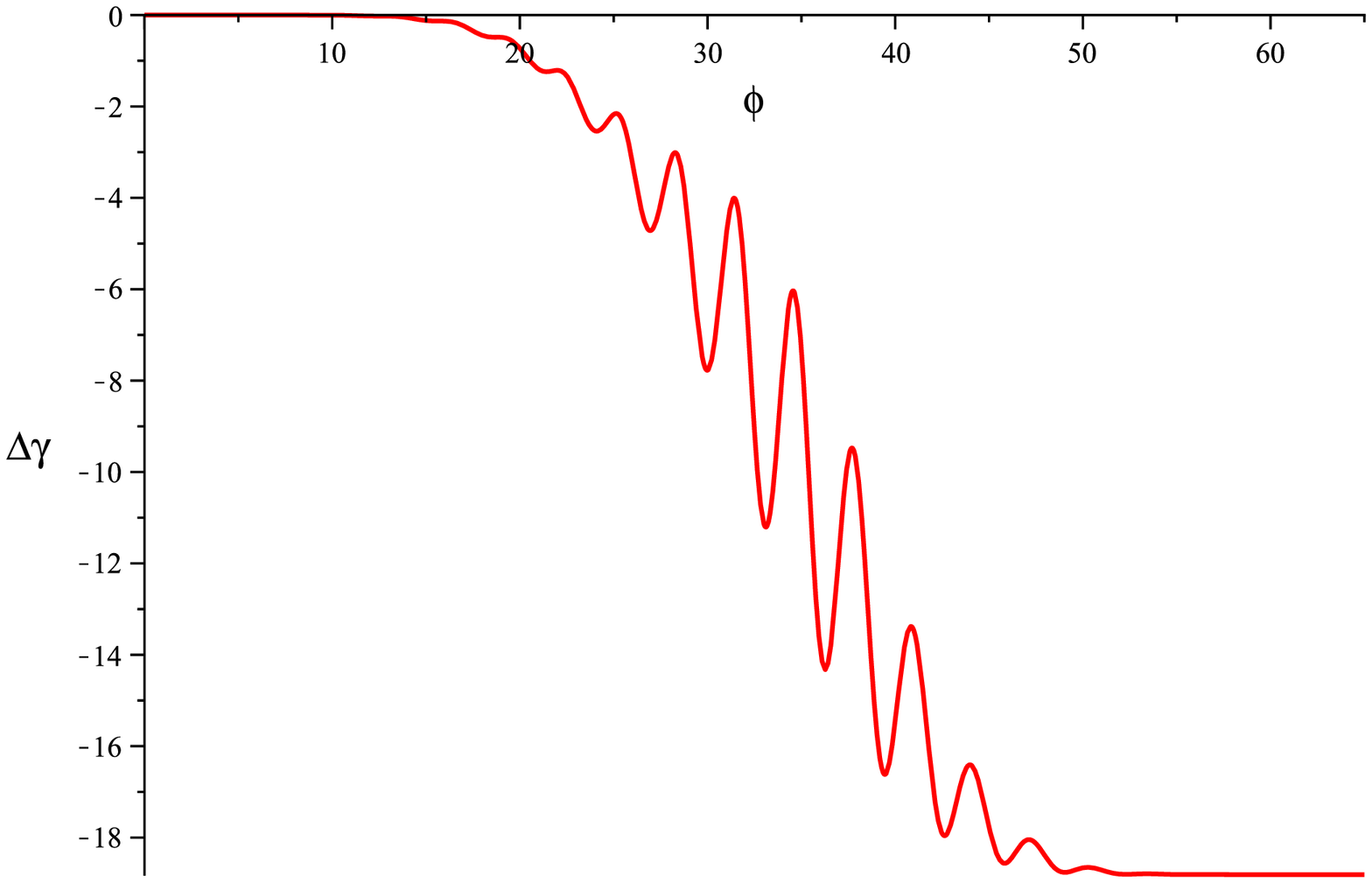}
\caption{\label{fig:DELTA.GAMMA} RR correction to electron gamma factor as a function of phase $\phi$ for $a_0 = 150$ but different values of the initial gamma factor. Upper panel: $\gamma_0 = 1$ (FTM). Central panel: $\gamma_0 = 40$. Lower panel: $\gamma_0 = 100$.}
\end{figure}

Let us finally trace the behaviour of the RR correction $\Delta \gamma$ during the crossover from FTM to CM, i.e.\ with increasing $\gamma_0$. This is depicted in Fig.~\ref{fig:DELTA.GAMMA}. The upper panel shows the FTM ($\gamma_0 = 1$). In agreement with (\ref{DELTA.FTM}) the RR correction is always positive so that there is energy gain within the pulse. Compared to the amplitude $\gamma$ (see Fig.~\ref{fig:GAMMA1}) this is rather small, $\Delta \gamma_\mathrm{FTM} \ll \gamma$. This intermediate energy gain increases with $a_0$, cf.\ (\ref{DELTA.FTM}), but no net energy transfer survives the passing of the pulse.

If $\gamma_0$ is increased, $\gamma_0 > 1$, one enters CM (Fig.\ \ref{fig:DELTA.GAMMA}, central and lower panel). For intermediate values of $\gamma_0$ (central panel), $\Delta \gamma_\mathrm{CM}$ stays positive as long as $\phi$ is not too large. Once a sufficient number of cycles has passed, however, $\Delta\gamma_\mathrm{CM}$ becomes negative which implies net energy loss.

For sufficiently large $\gamma_0$, the RR correction $\Delta\gamma_\mathrm{CM}$ is always negative, irrespective of the value of $\phi$ (Fig.\ \ref{fig:DELTA.GAMMA}, lower panel). In summary, Fig.\ \ref{fig:DELTA.GAMMA} thus shows the competition of the second and third terms in (\ref{DELTA.GAMMA.I}). The latter is absent for $\gamma_0 = 1$ (FTM, upper panel). For intermediate $\gamma_0$ (central panel) the $a_0^2$ term dominates for small $\phi$ but, since $I_2$ increases monotonically with $\phi$, it overwhelms the sum for large $\phi$ when $I_4$ goes to zero. For sufficiently large $\gamma_0$ this holds for all $\phi$ (lower panel).

This competition of $a_0$ and $\gamma_0$ has been noted before in the context of nonlinear Thomson scattering \cite{Harvey:2009ry} and also RR dynamics \cite{DiPiazza:2009zz}. There it was found that the 3-momentum transfer changes sign at a ``critical'' value of $a_0 \simeq 2 \gamma_0$. For Thomson scattering this value defines an effective centre-of-mass system for which there is no energy transfer between electrons and laser photons. As a result, the theoretical emission spectrum degenerates into a pure line spectrum \cite{Harvey:2009ry}.

We close this section be reemphasising the crucial importance of working consistently to leading order in RR. Painful experience has told us that (wrongly) including higher order terms in the LL solution (\ref{V.SOLUTION}) distorts the balance of terms in (\ref{DELTA.GAMMA.I}), and one would not reach the conclusions above.

\section{Conclusion}

We have re-analysed the problem of radiation reaction by solving the Landau-Lifshitz equation analytically for an electron in an intense plane wave laser field. Such a field depends solely on the phase $\phi = k \cdot x$ or, with the laser wave vector $\vcb{k}$ pointing in $z$ direction, on the light-front coordinate $x^\m = ct - z$. A particularly useful signature for radiation reaction is the laser frequency as `seen' by the electron, $\Omega = k \cdot u$. This ceases to be conserved when radiation reaction is present and thus provides a clear signal for symmetry breaking: translational invariance in the light-front coordinate $x^\p = ct + z$ is lost.

For a pulsed plane wave of finite duration in $x^\m$ (consisting of $N$ cycles) the magnitude of the total change in $\Omega$ (hence in longitudinal momentum, $p^\m$) obtained from the Landau-Lifshitz equation is well described by Larmor's formula for the radiated power. This has been corroborated by a careful study of the energy transfer dynamics and its dependence on initial conditions. The fixed target mode (electron at rest initially) is singled out as the only scenario for which there is no net energy transfer. Colliding modes with arbitrarily small initial velocity ($\gamma_0 > 1$) always entail net energy loss. To arrive at these results it is of crucial importance to consistently truncate all expressions at leading order in radiation reaction. In this way it becomes obvious that the Landau-Lifshitz equation breaks down when $\gamma_0 a_0^2 N \simeq 10^8$. Hence, if one wants to rely on the Landau-Lifshitz approximation, the electron energy $\gamma_0$, laser amplitude $a_0$ and pulse duration $N$ cannot be increased arbitrarily.

The integrability properties of both the Lorentz and Landau-Lifshitz equations seem quite intriguing. We plan to return to this topic elsewhere.

\acknowledgements

The authors acknowledge stimulating discussions with S.S.\ Bulanov, A.\ di Piazza and A.\ Ilderton. C.H.\ was supported by the Swedish Research Council Contract \#2007-4422 and the European Research Council Contract \#204059-QPQV

\bibliographystyle{apsrev4-1}

\begin{thebibliography}{99}

\bibitem{Lorentz:1892}
  H.A.~Lorentz,
  \textit{La Th{\'e}ori{\'e} Electromagn{\'e}tique de Maxwell et son Application aux Corps Mouvants},
  Arch.\ Ne{\'e}rl.\ \textbf{25}, 363-552 (1892), reprinted in Collected Papers (Martinus Nijhoff, The Hague, 1936), Vol. II, pp. 64-343.

\bibitem{Lorentz:1909}
  H.A.~Lorentz, \textit{The Theory of Electrons}, B.G. Teubner, Leipzig, 1906;
  reprinted by Dover Publications, New York, 1952 and Cosimo, New York, 2007.

\bibitem{Abraham:1905}
  M.~Abraham, \textit{Theorie der Elektrizit{\"a}t}, Teubner, Leipzig, 1905.

\bibitem{Dirac:1938}
  P.A.M. Dirac,
  Proc.\ Roy.\ Soc.\ A \textbf{167}, 148-169 (1938).

\bibitem{Spohn:2004ik}
  H.~Spohn, \textit{Dynamics of Charged Particles and their Radiation Field},
  Cambridge University Press, Cambridge, 2004

\bibitem{Rohrlich:2007}
  F.~Rohrlich, \textit{Classical Charged Particles}, 3rd ed., World Scientific, Singapore, 2007.

\bibitem{McDonald:1998}
  K.T.~McDonald,
  \textit{Limits on the Applicability of Classical Electromagnetic Fields as Inferred from the Radiation Reaction}, unpublished preprint, available from: \\ http://www.hep.princeton.edu/\~{}mcdonald/examples/

\bibitem{Landau:1987}
  L.~D.~Landau and E.~M.~Lifshitz, \textit{The Classical Theory of Fields
    (Course of Theoretical Physics, Vol.~2)}, Butterworth-Heinemann, Oxford,
  1987.

\bibitem{Russo:2009yd}
  J.~G.~Russo and P.~K.~Townsend,
  J.\ Phys.\ A  {\bf 42}, 445402 (2009)

\bibitem{Spohn:1999uf}
  H.~Spohn,
  \textit{The Critical manifold of the Lorentz-Dirac equation},
  Europhys.\ Lett.\  {\bf 49}, 287 (2000)

\bibitem{Gralla:2009md}
  S.~E.~Gralla, A.~I.~Harte and R.~M.~Wald,
  \textit{A Rigorous Derivation of Electromagnetic Self-force},
  Phys.\ Rev.\  D {\bf 80}, 024031 (2009)

\bibitem{Bamber:1999zt}
  C.~Bamber, S.~J.~Boege, T.~Koffas, T.~Kotseroglou, A.~C.~Melissinos, D.~D.~Meyerhofer, D.~A.~Reis, W.~Ragg {\it et al.},
  Phys.\ Rev.\  {\bf D60}, 092004 (1999).

\bibitem{Heinzl:2008rh}
  T.~Heinzl and A.~Ilderton,
  Opt.\ Commun.\  {\bf 282}, 1879 (2009)

  \bibitem{Koga:2005}
  J.~Koga, T.Zh.~Esirkepov and S.V.~Bulanov, Phys.\ Plasmas \textbf{12}, 093106 (2005).

\bibitem{DiPiazza:2008lm}
  A.~Di Piazza,
  Lett.\ Math.\ Phys.\ \textbf{83}, 305 (2008).

\bibitem{DiPiazza:2009pk}
  A.~Di Piazza, K.~Z.~Hatsagortsyan and C.~H.~Keitel,
  Phys.\ Rev.\ Lett.\  {\bf 102}, 254802 (2009)

\bibitem{Hadad:2010mt}
  Y.~Hadad, L.~Labun, J.~Rafelski, N.~Elkina, C.~Klier, H.~Ruhl,
  Phys.\ Rev.\  {\bf D82}, 096012 (2010).

\bibitem{Sauter:1931zz}
  F.~Sauter,
  Z.\ Phys.\  {\bf 69}, 742-764 (1931).

  \bibitem{Yanovsky:2008}
  V.~Yanovsky et al.,
  Opt.\ Express \textbf{16}, 2109 (2008).

  \bibitem{Vulcan10PW:2009}
  The Vulcan 10 Petawatt Project: \\ http://www.clf.rl.ac.uk/New+Initiatives/14764.aspx

  \bibitem{ELI:2009}
  The Extreme Light Infrastructure (ELI) project: http://www.extreme-light-infrastructure.eu

\bibitem{Heinzl:2000ht}
  T.~Heinzl,
  Lect.\ Notes Phys.\  {\bf 572}, 55-142 (2001).

\bibitem{Synge:1935zzb}
  J.~L.~Synge, University of Toronto Applied Mathematics Series, No.~1 (Univ.\ of Toronto Press, 1935)

\bibitem{Stephani:2004ud}
  H.~Stephani, \textit{Relativity: An Introduction to Special and General Relativity}, Cambridge University Press, Cambridge, 2004.

\bibitem{Becker:1991}
  W.~Becker,
  Laser and Particle Beams \textbf{9}, 603 (1991).

\bibitem{Davis:1979zz}
  L.~W.~Davis,
  Phys.\ Rev.\  {\bf A19}, 1177 (1979).

\bibitem{Harvey:2011}
  C.~Harvey and M.~Marklund,
  \textit{Radiation damping in pulsed Gaussian beams},
  to appear

  \bibitem{Lai:1980}
  H.M.~Lai,
  Phys.\ Fluids \textbf{23}, 2373 (1980).


\bibitem{Harvey:2010ns}
  C.~Harvey, T.~Heinzl, N.~Iji, K.~Langfeld,
  Phys.\ Rev.\  {\bf D83}, 076013 (2011).


\bibitem{Mackenroth:2010jk}
  F.~Mackenroth, A.~Di Piazza, C.~H.~Keitel,
  Phys.\ Rev.\ Lett.\  {\bf 105}, 063903 (2010).

\bibitem{Heinzl:2010vg}
  T.~Heinzl, A.~Ilderton, M.~Marklund,
  Phys.\ Lett.\  {\bf B692}, 250 (2010).

%
%

\bibitem{Harvey:2009ry}
  C.~Harvey, T.~Heinzl, A.~Ilderton,
  Phys.\ Rev.\  {\bf A79}, 063407 (2009).

\bibitem{DiPiazza:2009zz}
  A.~Di Piazza, K.~Z.~Hatsagortsyan, C.~H.~Keitel,
  Phys.\ Rev.\ Lett.\  {\bf 102}, 254802 (2009).

\end{thebibliography}

\end{document}